\documentclass{article}
\usepackage{epsfig}
\usepackage{amsmath}
\newcommand{\asb}{\bar{\alpha}_s}
\newcommand{\omp}{\omega_\mathrm{I\!P}}

\def\kbo{{\bf k}}

\def\rbo{{\bf r}}  
\def\rbop{{\bf r}^\prime}  
\def\bbo{{\bf b}}

\def\gev{\mbox{\rm GeV}}

\begin{document}

\vspace*{2cm}

\begin{center}  
{\Large \bf Saturation at low $x$ and nonlinear evolution}\footnote{{Talk given at the Epiphany Conference, Cracow, January 3-6, 2002.}  }\\  
\vspace*{0.4in}  
 A.\ M. \ Sta\'sto\\  
\vspace*{0.5cm}  
  {\it  INFN Sezione di Firenze,  Sesto Fiorentino (FI), Italy} \\  
\vskip 2mm  
and
\vskip 2mm  
{\it H. Niewodnicza\'nski Institute of Nuclear Physics,  
 Krak\'ow, Poland} \\  
\vskip 2mm  

\end{center}  
\vspace*{1cm}

\begin{abstract}
In this talk the results of the analytical and numerical analysis of the nonlinear Balitsky-Kovchegov equation are presented. The characteristic BFKL diffusion into infrared regime is suppressed by the generation of the saturation scale $Q_s$. We identify the scaling and linear regimes for the solution. 
We also study the impact of  subleading corrections onto the nonlinear evolution.
\end{abstract}

\section{Introduction}
One of the major challenges in QCD is the description of high energy scattering phenomena. In the high center-of-mass energy $\sqrt{s}$ and in the perturbative domain when $\alpha_s \ll 1 $ the scattering amplitude is obtained by the summation of diagrams leading in $\log s $ \cite{BFKL}. At this level, when $\alpha_s$
is frozen, the dependence of the resulting cross section on the energy is governed by the power law $x^{-\omp}$ where $x$ is the Bjorken variable. The critical exponent $\omp = 4 \, \ln 2\, \asb$ ($\asb = \alpha_s N_c / \pi$) is provided by the minimum of the eigenvalue function $\chi(\gamma)$ of the BFKL evolution kernel. 

The conceptual problem in this approach is the fact
 that at sufficiently high center-of-mass energies the BFKL amplitude
 violates the Froissart unitarity bound. This means that the validity of this approach is strongly limited and it has to be modified at very small $x$ in order to guarantee the unitarity of the resulting cross section.

The solution to this problem can be provided by including the recombination
effects, which are likely to occur at very small values of $x$ \cite{GLR}. 
By decreasing the value of $x$ at fixed gluon virtuality $k_t^2$,
the density of partons becomes so large that they start to overlap.
In this case the gluon splitting process must be supplemented by a competing
gluon recombination.  In terms of evolution the master equations become
nonlinear with an additional quadratic term which suppress the growth
of the amplitude with energy and restore the unitarity\footnote{Recently \cite{Kovner} it has been pointed out that the situation can be actually more complicated in a sense that the Balitsky-Kovchegov equation \cite{Kovchegov} could lead to the local saturation but not to the unitarisation due to the fact that the target radius in impact parameter space could grow as fast as a power with energy.}.

There have been extensive studies on this problem, see \cite{GLR}, \cite{MUELLERQIU}-\cite{Kovchegov}, 
which result in the nonlinear evolution. One of  the important outcomes
of these studies is  the existence of the saturation scale $Q_s(x)$ which
is a characteristic scale at which the parton recombination effects
become important. In particular case of the Balitsky-Kovchegov equation \cite{Kovchegov} the existence of such scale has yet another important
impact on the picture of the BFKL evolution. The diffusion into the infra-red,
which is  the charactersitic property of BFKL evolution is strongly
limited due to  the existence of the saturation scale \cite{BAL3,GBMS}.
 In fact, in the regime
when the gluon transverse momenta $k<Q_s(x)$, the solution to the nonlinear equation \cite{Kovchegov} becomes a function of only one combined variable $k/Q_s(x)$ \cite{BRAUN1}-\cite{BL}. 
In the regime of high momenta, $k>Q_s(x)$ the parton density is small and the evolution is governed by a normal linear equation.

In this talk we present the analytical and numerical analysis of the 
 Balitsky - Kovchegov \cite{Kovchegov} equation which is a nonlinear evolution equation in the leading $\log s $ limit. We illustrate the emergence of the saturation scale and scaling and show that it leads to the suppression of the infra-red diffusion. We also consider the case with additional NLL effects such as kinematical constraint and running coupling.

 The results presented in this talk have been obtained in the collaboration with K. Golec - Biernat and L. Motyka. For the details of the calculation the reader is referred to  \cite{GBMS}.

\section{Nonlinear evolution equation}

The Balitsky-Kovchegov equation \cite{Kovchegov} has been derived as an evolution equation
for the dipole-nucleus amplitude in the dipole picture by a summation
of multiple Pomeron exchanges in the leading $\log s$ level and in the large $N_c$ limit.

The resulting evolution equation reads

\begin{multline} 
\label{eq:kov}  
 \frac{\partial N(\rbo,\bbo,Y)}{\partial Y}   \,=\, 
\overline{\alpha}_s\  
(K\otimes N)(\rbo,\bbo,Y)  - \\  
  -    \overline{\alpha}_s\,  
\int \frac{d^2\rbop}{2\pi}\,  
\frac{r^2}{r^{\/\prime 2}(\rbo+\rbop)^2}\, 
N(\rbo+\rbop,\bbo+\frac{\rbop}{2},Y)\;N(\rbop,\bbo+\frac{\rbo+\rbop}{2},Y) \; ,
\end{multline} 
where $\overline{\alpha}_s=N_c \alpha_s/\pi$, and the linear term is 
determined by the BFKL kernel  
\begin{multline} 
\label{eq:bfklkernel}  
(K\otimes N)(\rbo,\bbo,Y)= \\   
=\int \frac{d^2\rbop}{\pi\/ r^{\/\prime 2}}\,  
\left\{  
\frac{r^2}{(\rbo+\rbop)^2}\,N(\rbo+\rbop,\bbo+\frac{\rbop}{2},Y)  
-  
\frac{r^2}{r^{\/\prime 2}+(\rbo+\rbop)^2}\,N(\rbo,\bbo,Y)  
\right\}. 
\end{multline}  
The function $N(\rbo,\bbo,Y)$ is the dipole-nucleus amplitude for the 
scattering of the dipole with transverse size $\rbo$ at impact parameter $\bbo$
and at rapidity $Y$. 

In the linear approximation, when each dipole scatters 
only once off  
the nucleus, the BFKL equation in the dipole picture is obtained.  
The non-linear term in (\ref{eq:kov}) takes into account multiple scatterings  
and is essentially determined by the triple pomeron vertex \cite{BARWUE}  
in the large $N_c$ limit.  
Eq.~(\ref{eq:kov}) unitarizes the BFKL pomeron in the sense that at  
$x\rightarrow 0$ and $Q^2$ fixed,  
\begin{equation}  
F_2\, \sim\, Q^2\, \ln(1/x)\,.  
\end{equation} 
Thus the power-like  
rise with energy for the BFKL pomeron is tamed \cite{KOV2}.

For the subsequent analysis we shall assume the approximation of the big nucleus, i.e. $\rbo \ll \bbo$ which allows us to factorize the impact parameter dependence in Eq.(\ref{eq:kov}). We also consider the spherical symmetric solutions in $\rbo$ and transform the equation (\ref{eq:kov}) into the momentum space by performing the Fourier transform
\begin{equation}  
\phi(k,Y)  
\,=\, 
\int \frac{d^2\rbo}{2\pi} \exp(-i\kbo\cdot \rbo)\,  
\frac{N(r,Y)}{r^2}  
\,=\,  
\int_0^\infty  
\frac{dr}{r}\, J_0(k\/r)\,N(r,Y),  
\end{equation} 
where $J_0$ is the Bessel function.  
In this case the following equation  
is obtained  
\begin{align} 
\label{eq:newkov}  
\frac{\partial \phi(k,Y)}{\partial Y}  
\,=\,  
\overline{\alpha}_s\, (K^\prime\otimes \phi)(k,Y)  
\,-\  
\overline{\alpha}_s\, \phi^{\,2}(k,Y), 
\end{align} 
and the action of the BFKL kernel  is given by  
\begin{equation}  
(K^\prime\otimes \phi)(k,Y)  
\,=\,  
\int_0^{\infty} \frac{dk^{\prime 2}}{k^{\prime 2}}\,  
\left\{  
\frac{k^{\prime 2}\,\phi(k^{\prime},Y)\, -\, k^2\, \phi(k,Y)}  
{|k^2\,-\, k^{\prime 2}|}  
\,+\,  
\frac{k^2\, \phi(k,Y)}{\sqrt{4 k^{\prime 4}\,+\,k^4}}  
\right\} ,  
\end{equation}  
where now $k$ and $k^{\prime}$ are the transverse momenta of the exchanged gluons 
in the BFKL ladder.  
\section{Saturation scale and geometric scaling}

\begin{figure}  
  \vspace*{0.0cm}  
     \centerline{  
         \epsfig{figure=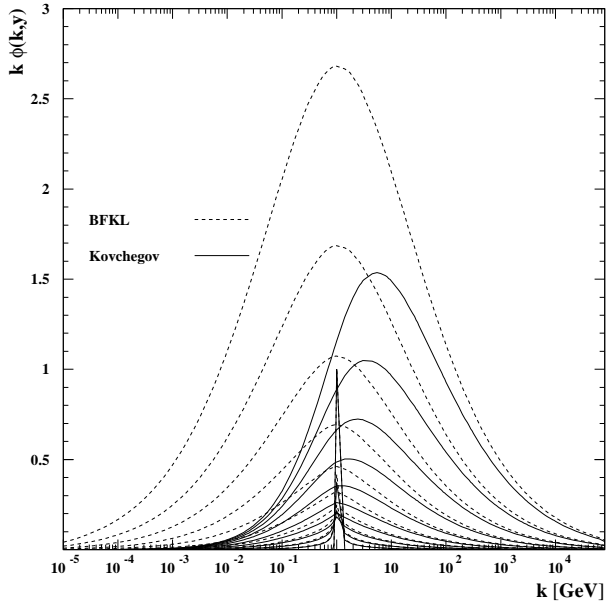,width=9cm}  
           }  
\vspace*{0.5cm}  
\caption{\it  
The functions $k\phi(k,Y)$ constructed from solutions to the BFKL and  
the Balitsky-Kovchegov equations with the delta-like input   
for different values of the evolution parameter $Y=\ln(1/x)$ ranging from $1$ to $10$.  
The coupling constant $\alpha_s=0.2$. 
\label{fig:distr}}  
\end{figure}

In order to study the diffusion and scaling properties of the solution
we shall consider the function $k \phi(k,Y)$. In the case of the linear BFKL
equation this function is a Gaussian concentrated around some initial scale $k_0$ and with  width increasing with rapidity, leading to a diffusion.
In Fig.\ref{fig:distr} we illustrate this distribution for the case of the solution to the linear BFKL and in the nonlinear Balitsky-Kovchegov equation for different choices of rapidity. As an initial condition we have chosen a delta function $\delta(k-k_0)$. In the case of the linear BFKL evolution the solution
is always peaked at $k=k_0$ and exhibits the well known diffusion pattern, in which the momentum distribution
has an increasing width with growing rapidity. In the nonlinear case however the solution behavies quite differently. The peak of the distribution $k_{\mathrm{max}}$ moves towards higher $k$ values as the rapidity increases and the solution becomes washed out from the 
$k<k_0$ regime. Only in the  region $k \gg k_{\mathrm{max}}$ it coincides with the linear evolution. 

\begin{figure}  
  \vspace*{0.0cm}  
     \centerline{  
         \epsfig{figure=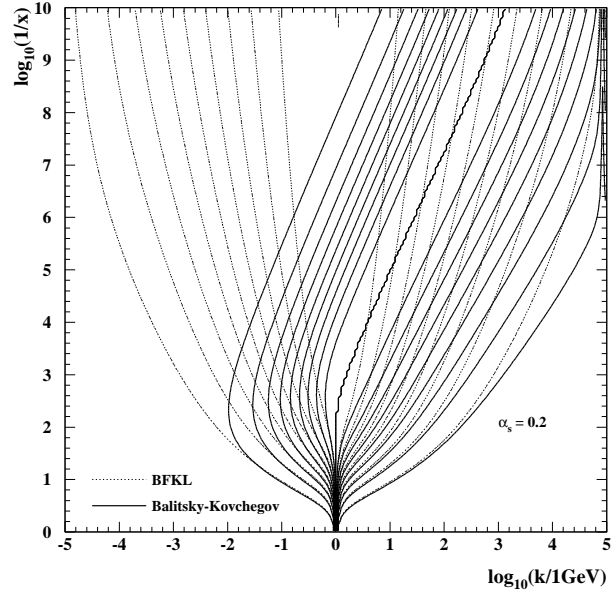,width=9cm}  
           }  
\vspace*{0.5cm}  
\caption{\it The lines of constant values of the BFKL and the BK  
re-normalized solutions  $\Psi(k,Y)$ ($Y=\ln (1/x)$) in the  
$(\log_{10}(k),\log_{10}(1/x))$-plane.  
\label{fig:proj}}  
\end{figure}

The impact of unitarization of the BFKL pomeron on the infra-red behaviour  
can be also visualised by studying the properties of the following normalised distribution 
\begin{equation}  
\label{eq:norm}  
\Psi(k,Y)\,=\,  
\frac{k\,\phi(k,Y)}  
{k_{\mathrm{max}}(Y)\,\phi(k_{\mathrm{max}}(Y),Y)},  
\end{equation}  
and by performing the projection of this function onto the ($\ln k/k_0,Y$)
plane, Fig.~\ref{fig:proj}.

Again,  
for small $Y$, when the non-linearity in the BK equation is negligible,  
the re-normalized solutions (\ref{eq:norm})
 of the BFKL and the BK equations  coincide.  
With  increasing Y,  when the  
non-linear effects become important, the difference between them  
in the region of small $k$  becomes fully visible.  
Note that in certain region of  ($\ln k/k_0,Y$) in  Fig.~\ref{fig:proj}
the contours become parallel straight lines.
This means that $\Psi(k,Y)$ in this region is a function  of the  
combination  
\begin{equation}  
\label{eq:scalvar} 
\xi\,=\,  
\ln(k/k_0)\,-\,\lambda\,Y 
\,=\, \ln\left(\frac{k}{k_0 \exp(\lambda Y)} \right)~~~~~~~~~~~~~\lambda>0  \; ,
\end{equation}  
 instead of $k$  and $Y$ separately. This is referred as the geometric scaling property.
We note, that this scaling  holds in the regime where $k<k_{\mathrm{max}}$ and becomes violated when $k>k_{\mathrm{max}}$. This suggests that we can identify the saturation scale $Q_s(Y)$ with $k_{\mathrm{max}}$
\begin{equation}  
\label{eq:satscale}  
Q_s(Y)\equiv k_{\mathrm{max}}(Y)\,=\,Q_0 \exp(\lambda Y)\,,  
\end{equation}  
with the exponential  
dependence on rapidity governed by the value of the scaling parameter 
$\lambda$.  

The solution to Eq.~(\ref{eq:newkov})  
has the same scaling property as the function $\psi(k,Y)$ namely 
\begin{equation}  
\phi(k,Y)=\phi(k/Q_s(Y)) \; , 
\label{eq:kmaxscal}  
\end{equation}  
provided that $\phi(k_{max},Y)=\mathrm{const}$. We have checked that this condition is satisfied.

We have checked that the scaling coefficient $\lambda$ (\ref{eq:satscale}) is a universal quantity and it does not depend on the type of the input distribution whereas the normalisation $Q_0$ is determined by the initial conditions.

\begin{figure}
  \vspace*{0.0cm}  
     \centerline{  
         \epsfig{figure=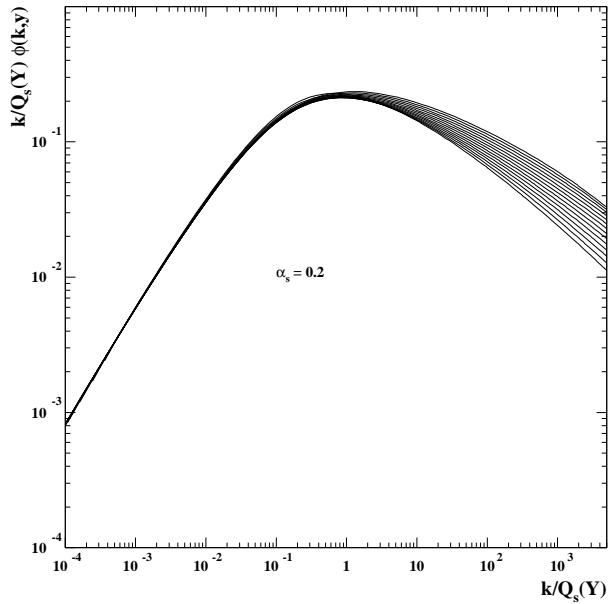,width=9cm} 
           }  
\vspace*{0.5cm}  
\caption{  
The function $(k/Q_s(Y))\,\phi(k,Y)$  
plotted versus $k/Q_s(Y)$ for different values of rapidity $Y$  
ranging from $10$ to $23$. The saturation scale $Q_s(Y)$ corresponds to the position 
of the maximum of the function $k\,\phi(k,Y)$.  
\label{fig:trans}}  
\end{figure}  
The transition between the scaling and linear regime can be perhaps best seen  in Fig.~\ref{fig:trans} where  
the function $k/Q_s(Y)\,\phi(k,Y)$ is plotted as a function of the scaling  
variable $k/Q_s(Y)$ for different values of the rapidity $Y$. 
The scaling behaviour is 
represented by a common line to the left of the point $k/Q_s(Y)=1$. The slow departure  
from scaling for $k>Q_s(Y)$ is clearly visible.  
In this sense the line $k=Q_s(Y)$ is  
only an approximation characterizing the position of the transition region  
in the $(k,Y)$-plane. However, the choice based on the position in $k$ of the  
maximum of $k\phi(k,Y)$ as a function of $Y$ is the most natural one.

From the numerical solution we have extracted the value of the scaling parameter and found $\lambda \simeq 2.05 \, \asb$ which is in agreement with the estimates of \cite{LEVTUCH1,LEVTUCH2} for the Balitsky-Kovchegov equation and also with the previous work \cite{BL}.

The geometric scaling is also the property of the Golec-Biernat and W\"usthoff saturation model \cite{GBW1} which succesfully described the data on inclusive and diffractive proton structure function. In this model it is assumed that the whole energy dependence of the dipole cross section $\sigma_d(r,Y) = \sigma_d(r^2 Q_s^2(Y))$ is driven through the saturation scale $Q_s(Y)$ (or saturation radius $1/Q_s(Y)$). It then results in an approximate scaling property of the data for the total photon-proton cross section 
\begin{equation}  
\sigma_{\gamma^*p}(x,Q^2)\,=\,\sigma_{\gamma^*p}(Q^2/Q_s^2(Y)).  
\end{equation}  
Such  scaling law  was found in  
the small-$x$ DIS data \cite{SGBK}.  
\section{Analysis beyond LL level}

The Balitsky-Kovchegov equation has been derived in the leading $\log s$
level, therefore it is important to study the impact of the subleading corrections. Although it has been argued \cite{Unitarity} that the unitarity corrections might become important before next-to-leading ones, the study of the latter ones is crucial due to the large numerical value  of the subleading series.
After the evaluation of the NLL contribution to the BFKL amplitude has been performed \cite{NLLBFKL} it turned out that the correction is very large and need to be resummed \cite{RESUM} in order to stablize the final result.

We have tested two important types of the NLL corrections: the inclusion of running of the coupling $\asb(k^2)$ and the so called kinematical constraint.

In the case of linear BFKL the running of the coupling poses serious problems due to the existence of the Landau pole and thus it is  necessary  to regularise  $\asb(k^2)$ at small scales. This results in a strong dependence of the solution on the cut-off (or freezing ) parameter $k_0$.
\begin{figure}[t]
  \vspace*{0.0cm}  
     \centerline{  
         \epsfig{figure=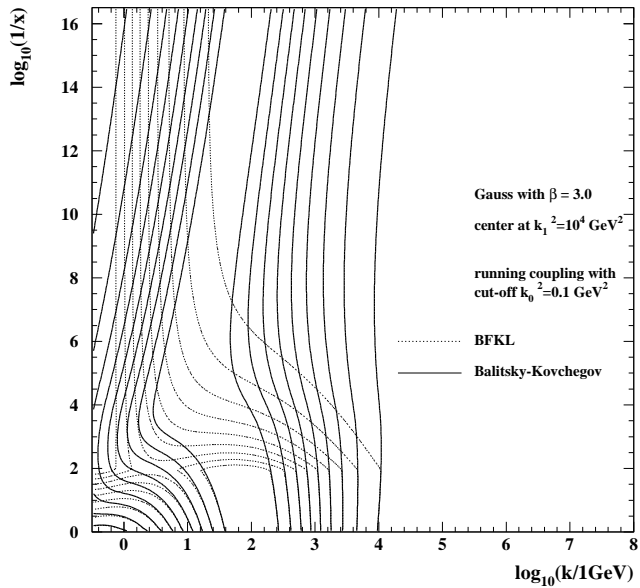,width=9cm}  
           }  

\caption{\it  
The re-normalized solution  $\Psi(k,Y)$  
for 
the Balitsky-Kovchegov equation with the running coupling constant  
and the infra-red cut-off $k_0^2 = 0.1\, \gev^2$ 
as function of  $\log_{10}(1/x)$ and $\log_{10}(k/1\,\gev)$.  
\label{fig:conrun}}  
\end{figure}  
The intercept of the BFKL Pomeron turns out to be dominated by the values of 
$\asb(k_0^2)$ and instead of the typical diffusion pattern one has a factorized behaviour of the solution $k \phi (k,Y)\,\sim\,  
\exp\{\lambda Y\}\, {\frac{1}{k}}\,[\ln \left({k^2}/{k_0^2}\right)]^{\nu}$ at large rapidities.
The distribution of the gluon momenta is dominanted by the virtualities in the infrared regime $\sim k_0$. 

In the case of the Balitsky-Kovchegov equation with the running coupling 
one has still problem of regularisation around the Landau pole, however the solution
is itself much more stable with respect to the details of the phenomenological regularisation. It turns out, that since the saturation effects are very strong
in the regime of small values of $k$, they tend to decrease the rapid rise of the amplitude for the values of running coupling evaluated at scales around the cutoff scale $k_0$.
The rapidity - dependent saturation scale, $Q_s(Y)$ shifts the momentum distribution
out of the infrared regime into the perturbative domain.
This phenomenon can be best visualized by means of similar contour plots as before, Fig.\ref{fig:conrun}.
In the linear BFKL case one observes that the solution initially undergoes the diffusion pattern, slightly asymmetric due to corrections coming from the  running coupling, and then suddenly it drops into the small scales regime $k \sim k_0$ where it exhibits factorized (in $Y$ and $\ln k/k_0$) behaviour. On the contrary, the Balitsky -Kovchegov equation which initially coincides with BFKL, also undergoes some form of  transition  but then the distribution moves away from the infrared regime due to the generation of the saturation scale. 
\begin{figure}
  \vspace*{0.0cm}  
     \centerline{  
         \epsfig{figure=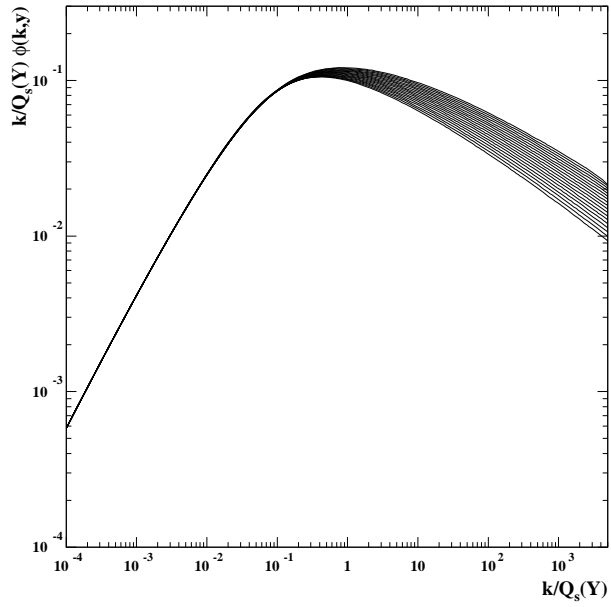,width=9cm}  
           }  
\vspace*{0.5cm}  
\caption{\it  
The function $(k/Q_s(Y))\,\phi(k,Y)$ in the  running coupling  case,  
plotted versus $k/Q_s(Y)$ for different values of rapidity $Y$  
ranging from $15$ to $32$. The saturation scale $Q_s(Y)$ is taken from equation  
(\ref{eq:satscalerc}) with the initial condition $Q_s(Y=0)=2.0 \; GeV$.  
\label{fig:scalerun}}  
\end{figure}  
We have estimated the saturation scale in the case of Balitsky-Kovchegov equation with running coupling.
By taking the  ansatz that 
the local exponent $\lambda(Y)= d \ln(Q_s(Y)/\Lambda)\; /dY$  
takes the form $\lambda(Y) = 2 \bar\alpha_s (Q^2_s(Y))$ where $\Lambda=\Lambda_{QCD}$
we have found that the saturation scale is given
by the solution to the differential equation
\begin{equation}  
{d\ln (Q_s(Y)/\Lambda) \over dY} = {12 \over b_0 \ln ( Q_s(Y) /\Lambda)}\,,  
\end{equation} 
with the initial condition $Q_s(Y_0) = Q_0$  and  $Y_0$ chosen in the region where 
scaling sets in. The solution takes the form  
\begin{equation}  
Q_s(Y) = \Lambda\; \exp\left( \sqrt{{24\over b_0}\, (Y-Y_0) + L_0 ^2} \right)\,,  
\qquad Y>Y_0\,,  
\label{eq:satscalerc} 
\end{equation}  
where $L_0 = \ln (Q_0/\Lambda)$.  
It follows that the local exponent $\lambda(Y)$ decreases with  
increasing rapidity, and $\lambda(Y) \sim 1/\sqrt{Y}$  for very large $Y$.  

Such dependence is indeed seen in the numerical analysis. 
 
In Fig. \ref{fig:scalerun} we illustrate scaling in the case with running coupling  by 
showing the function $(k/Q_s(Y)) \, \phi(k,Y)$ plotted versus 
$k/Q_s(Y)$ for different values of rapidity.  $Q_s(Y)$ is given  by 
formula (\ref{eq:satscalerc}) with the initial condition 
$Q_s(Y=0) = 2~\mbox{\rm GeV}$. 
The overlapping curves at low values of the scaling variable  
clearly indicate 
that for $k<Q_s(Y)$  scaling  is satisfied to a very good accuracy, 
thus justifying our ansatz (\ref{eq:satscalerc}) for the saturation  
scale. 
We have also tested the impact of the so called kinematical constraint onto the solution of Balitsky-Kovchegov equation and found that though qualitative features are the same (with generation of the saturation scale and scaling) the absolute numerical value of the solution is strongly decreased and also the saturation scale is shifted towards smaller value of gluon momenta.

To summarize, 
the diffusion into infrared in the case of 
the Balitsky-Kovchegov equation  is strongly suppressed due to the emergence of the saturation scale which can be approximately identified with the maximum of the momentum distribution $k \phi(k,Y)$. The solution to the nonlinear equation has  the property of the geometric scaling in the regime where $k<Q_s(Y)$ whereas in the case when $k>Q_s(Y)$ the solution enters the linear regime.
In the case of  running coupling 
the nonlinear equation  
 turns out to be more stable as compared with the linear BFKL evolution.
The sensitivity to the treatment of the infra-red region is much  
smaller than in the linear case due to the appearance of the saturation  
scale. 
In contrast to the  BFKL equation with the running coupling, the large $Y$  
asymptotics of the dipole scattering amplitude arising from the BK 
equation is governed by gluon momenta  
in the perturbative domain.

\section*{Acknowledgments} 
This research was supported in part by the EU Framework TMR  
programme, contract FMRX-CT98-0194 and by the Polish Committee for Scientific Research  
grants Nos.\ KBN~2P03B~120~19, 2P03B~051~19, 5P03B~144~20.


\end{document}